\documentstyle[12pt]{article}

\newcommand{\lesssim}{\mathrel{\raisebox{- .8 ex}
     {$\stackrel{\textstyle <}{\sim}$}}}

\begin{document}


\title{Continuous fuzzy measurement of energy for a two-level 
system}

\author{J\"urgen Audretsch
\and
Michael Mensky\thanks{Permanent address: 
        P.N.Lebedev Physical Institute, 117924 Moscow, Russia}\\
        Fakult\"at f\"ur Physik der Universit\"at Konstanz\\
        Postfach 5560 M 674, D-78434 Konstanz, Germany}

\date{November 25, 1996}

\maketitle

\begin{abstract}
A continuous measurement of energy which is sharp 
(perfect) leads to the quantum Zeno effect (freezing of the state). 
Only if the quantum measurement is fuzzy, continuous monitoring gives 
a readout $E(t)$ from which information about the dynamical 
development of the state vector of the system may be obtained
in certain cases. This is studied in detail. Fuzziness is thereby 
introduced with the help of restricted path integrals equivalent to  
non-Hermitian Hamiltonians. For an otherwise undisturbed multilevel 
system it is shown that this measurement represents a model of 
decoherence. If it lasts long enough, the measurement readout 
discriminates between the energy levels and the von Neumann 
state reduction is obtained. For a two-level system under resonance
influence (which undergoes in absence of measurement Rabi oscillations
between the levels) different regimes of measurement are specified
depending on its duration and fuzziness: 1) the Zeno regime where 
the measurement results in a freezing of the transitions between the
levels and 2) the Rabi regime when the transitions maintain. It is shown
that in the Rabi regime at the border to the Zeno regime a 
correlation exists between the time dependent measurement readout
and the modified Rabi oscillations of the state of the measured system. 
Possible realizations of continuous fuzzy measurements of energy 
are sketched. 
\end{abstract}

PACS: 03.65 Bz


\section{Introduction}\label{intro}

Rabi flopping is a well known phenomenon: If a two-level atom 
with energy levels $E_1$ and $E_2$ is under the action of a 
monochromatic electromagnetic field, which we assume for 
simplicity to be resonant, then the state vector of the atom 
oscillates between the lower and upper state with the Rabi 
frequency which depends on the amplitude of the driving field. 
Speaking more precisely, the probability for the atom to be, say, 
on the level $E_1$ oscillates sinusoidally between 1 and 0. 
Whenever a von Neumann (``instantaneous" and perfect) 
measurement of energy 
is made at a certain time, the measurement result is either $E_1$ 
or $E_2$ with the probability depending on the time when the 
measurement is performed, in accordance with this sinusoid. The 
atom is after the measurement in the energy eigenstate 
corresponding to $E_1$ or $E_2$. This means that such a 
measurement disturbs the Rabi oscillation radically: the phase of 
the Rabi flopping during next period is determined by the result 
of the von Neumann measurement. 

It is also well known that if perfect measurements of energy of the 
atom (von Neumann measurements) are very frequent and in the 
limit continuous, the interaction with the measuring apparatus 
prevents atomic transitions between levels. The atom stays in an 
eigenstate and the readout of the continuous perfect measurement 
reduces to the constant value $E_1$ or $E_2$. This is the often 
discussed quantum Zeno effect \cite{Zeno} which has been verified 
experimentally \cite{Itano}. It is impossible in this regime of 
measurement to obtain information about the Rabi oscillation. 
Moreover, the Rabi oscillation is prevented by the measurement 
because the influence of the measurement on the atom is stronger 
than the influence of the driving field. 

In contrast to this we want to address in the following the 
fundamental question if it is nevertheless possible to make the Rabi 
oscillation ``visible" by {\em continuous} measurement of the atomic 
energy, i.e. to obtain a correlation between the measurement readout 
$E(t)$ and the Rabi oscillation of the state vector. To achieve 
our goal we must first of all avoid the Zeno effect by weakening 
the influence of the measurement. This can be done with the help 
of continuous measurements which are fuzzy (unsharp) instead of 
being perfect (sharp). 

A {\em fuzzy} (unsharp) measurement is caused by the low 
resolution of the respective detector. In the case of an 
observable with a discrete spectrum fuzziness means that the 
measurement is not able to transfer the state of the system to an 
eigenvector. Fuzziness does not go back to a defect detector, but 
to its finite resolution. Note that in practice a fuzzy 
measurement is the generic case and that a sharp or perfect 
measurement may be considered only as an idealization. The study 
of continuous fuzzy measurements opens a very interesting new 
field in which fundamental problems of quantum theory as well as 
potentially important concrete applications are discussed. There 
are physical situations in which we have to rely on fuzzy 
measurements if we want to obtain relevant information 
\cite{Peres-book}. 

The theory of continuous quantum measurements has a long history. 
An incomplete list of references is
\cite{Zeh,cont-meas-count,M79,Barch,Peres-SG,%
Milburn-en,MilburnGa,Milburn,JoosZeh,Diosi,KMNamiot}.
The majority of authors studying continuous quantum measurements 
base their investigations on particular micro-models for the 
measuring device, see 
\cite{Peres-SG,Milburn-en,MilburnGa,Peres-book} for the 
measurement of the energy of a two-level system. As compared to 
this the phenomenological and therefore model-independent 
approach presented below has the big advantage that 
the universal structures of the theory of continuous measurements 
can be revealed, including those connected with fuzziness. This 
approach based on restricted path integrals (RPI). It has been  
proposed in \cite{M79,book93}. The approach has already been 
applied to the measurement of energy in \cite{OnoPre93,OnoPre95}. 
However, an {\em ad hoc} assumption was made in 
\cite{OnoPre93,OnoPre95} that the measurement readout is a 
constant curve $E(t)=E_n={\rm const}$. This is too restrictive and 
misleading for a continuous fuzzy measurement. 

A theory of continuous fuzzy measurements (say, monitoring 
some observable has to answer the following questions: 
How can the type and strength of fuzziness be represented 
theoretically and what will be the characteristic parameters? 
Let the measurement of the observable $A(p,q,t)$ be of 
duration $T$, what is the probability to obtain the 
measurement readout $[a] = \{a(t)|0\le t \le T\}$? If for an 
initial state $|\psi_0\rangle$ of a system a particular 
measurement readout $[a]$ is obtained, what will be the state  
$|\psi^{[a]}_T\rangle$ of the system at the end of the 
measurement? The way to answer these questions in the framework 
of the phenomenological RPI approach will be presented in 
Sect.~\ref{RPI}. The resulting dynamics is governed by a complex 
Hamiltonian depending on $[a]$. The probability for $[a]$ can be 
read off from $|\psi^{[a]}_T\rangle$. We obtain as a 
characteristic property of a continuous fuzzy measurement that 
its resolution improves with increasing duration $T$ as  $\Delta 
a_T \sim 1/\sqrt{T}$. This reflects a Zeno-type influence which 
is still present even if it does not result in the proper Zeno 
effect.\footnote{This phenomenon is referred to as ``narrowing" by 
Milburn in the second of Ref.~\cite{Milburn-en}.}

In Sect.~\ref{free} the scheme is applied to the measurement of 
energy of a multilevel system which is not influenced otherwise 
(no additional driving field). Different durations $T$ and the 
corresponding 
regimes of measurement are discussed. The time after which 
neighboring levels are resolved is called the level resolution 
time $T_{\rm lr}$. Its shortness is a measure for the strength of 
a Zeno-type influence (``level freezing") present in the 
continuous fuzzy measurement. The relation to the process of 
decoherence leading to the perfect (von Neumann) energy 
measurement is studied. 

A two-level system under the resonant influence of a periodical 
driving field is then investigated in detail in Sect.~\ref{two-level}. 
The driving field acts in opposition to the Zeno-type influence 
of the measurement lasting for the time $T$. Short Rabi periods 
$T_{\rm R}\sim$(Rabi frequency)$^{-1}$ represent strong 
influences of the driving field. It is obvious that different 
relative magnitudes of the times $T_{\rm lr}$, $T_{\rm R}$ and 
$T$ characterize different regimes of measurement with different 
behavior of the atom state. A complex general view is to be 
expected. We show that there is a regime in which a correlation 
exists between the time dependent measurement readout $E(t)$ and 
the Rabi oscillations of the atom state under the influence of 
the periodic driving field. 

In Sect.~\ref{realize} possible schemes of realization of the 
measurement in question are discussed. A continuous fuzzy 
measurement may be realized, according to these schemes, as a 
series of ``instantaneous" (short-time) unsharp measurements. 
Finally in Sect.~\ref{conclus} concluding remarks are made.

\section{A phenomenological approach to continuous fuzzy 
measurements}\label{RPI}

\subsection{Restricted path integrals}

Evolution of a quantum system when no measurement is 
performed (and therefore the system is closed)
can be described by a unitary evolution operator $U_t $.  
The matrix element $\langle q'' | U_t |q'\rangle = U_t(q'', q')$
of the evolution operator between states with definite 
coordinates (called the propagator) may be constructed with the 
help of the Feynman path integral (for example in the phase-space 
representation, i.e. with integration over paths $[p,q]$ in the 
phase space). 

The formalism of path integrals is physically transparent and can 
be generalized to different non-standard situations. One of them 
is the situation when the system undergoes a {\em continuous 
measurement} (and is therefore open).\footnote{Of course, the 
measurement must be precise enough so that quantum effects be 
essential; then it is called a quantum measurement.} The idea to 
apply {\em restricted} path integrals (RPI) in this case was 
suggested by R.Feynman \cite{Feynman48} and technically 
elaborated by one of the present authors (see \cite{M79} for the 
original papers and \cite{book93} for modern formulations).

A continuous measurement performed in the time interval $[0,T]$ 
gives some information about which paths the system propagates 
along. The central idea of the RPI approach is that the original 
path integral must be changed in accordance 
with this information. In the simplest case the path 
integral has to be restricted to a subset of paths depending on 
the measurement readout. In a more general case the contributing 
paths should be weighted by a weight functional depending on 
this readout. 

To be more concrete, let us consider as a typical continuous 
fuzzy measurement the monitoring of an observable $A(p,q,t)$ 
during the time interval $[0,T]$. In this case the registered 
{\em measurement readout} (output) is a curve
\begin{equation}
[a] = \{a(t)|0\le t \le T\}.
\end{equation}
Then the path integral has the form 
\begin{equation}\label{part-prop}
U_T^{[a]}(q'',q')
=\int d[p]d[q]\,w_{[a]}[p,q]\,
e^{\frac{i}{\hbar}\int_0^T (p\dot q - H(p,q,t))}.
\end{equation}
The weight functional $w_{[a]}$ will be discussed below. 

Evolution of the system undergoing continuous fuzzy measurement 
resulting in a certain readout $[a]$ may now be represented by the 
following formulas:
\begin{equation}\label{evolut}
|\psi_T^{[a]}\rangle = U_T^{[a]} |\psi_0\rangle, \quad
\rho_T^{[a]} = U_T^{[a]} \rho_0 \left(
U_T^{[a]}\right)^{\dagger}.
\end{equation}
These formulas contain  {\em partial propagators} depending on 
$[a]$. The first of these formulas may be applied to the 
evolution of pure states while the second is valid both for pure 
and mixed states. Note however that both of them are applicable 
only in the case when the measurement readout is known, i.e. in 
the situation of a {\em selective measurement}. A pure state 
remains pure after the selective measurement.

The norm of the state resulting from the measurement is, 
according to these formulas, less than unity. Normalization may 
of course be performed, but it is much more convenient and 
transparent to work  - as we will do - with non-normalized 
states. Their norms give according to \cite{book93} the 
{\em probability 
densities} $P[a]= {\rm Tr}\, \rho_T^{[a]}$ of the corresponding 
measurement readouts $[a]$. The following formula (with an 
appropriate measure of integration):
\begin{equation}\label{prob}
{\rm {Prob}}([a]\in{\cal A})=
\int_{{\cal A}} d[a]\, P[a]=
\int_{{\cal A}} d[a]\, {\rm Tr} \left( U_T^{[a]}
\rho_0 \left( U_T^{[a]}\right)^{\dagger}\right).
\end{equation}
gives the probability for a measurement to result in an readout 
from the set $\cal A$. The word `density' refers to the space of 
all measurement readouts.

In the considerations above we discussed the situation when 
the measurement readout is known (selective description 
of the measurement). If it is only known that the continuous 
fuzzy measurement is performed 
but its concrete readout is unknown (for example in {\em a 
priori} calculations), we have the situation of a {\em 
non-selective measurement}. In this case the evolution is 
described by the second of the formulas (\ref{evolut}) with 
summation over all possible measurement readouts:
\begin{equation}\label{nonselect-evol}
\rho_T = \int d[a]\,  \rho_T^{[a]}
=\int d[a]\, U_T^{[a]} \rho_0 \left( U_T^{[a]}\right)^{\dagger}.
\end{equation}
The density matrix $\rho_T$ is normalized for an arbitrary 
initial (normalized) state $\rho_0$ if the {\em generalized 
unitarity condition}
\begin{equation}\label{unitar}
\int d[a]\,  \left( U_T^{[a]}\right)^{\dagger}\, U_T^{[a]} =
\bf 1
\end{equation}
is valid.

In the most general case we have only a partial information about 
the measurement readout. If we know that $[a]\in{\cal 
A}$, the evolution of the system must be described by the 
formula
\begin{equation}\label{part-nonsel-evolut}
\rho_T^{{\cal A}} =
\int_{{\cal A}} d[a]\,  \rho_T^{[a]}
=\int_{{\cal A}} d[a]\,
U_T^{[a]} \rho_0 \left( U_T^{[a]}\right)^{\dagger}
\end{equation}
intermediate between Eq.~(\ref{evolut}) and 
Eq.~(\ref{nonselect-evol}).

To specify the type of fuzziness we are dealing with, we have to 
specify the weight functional $w_{[a]}$ in (\ref{part-prop}). To 
do so, we introduce the concept of closeness in the space of 
curves $[a]$ characterizing readouts of the continuous 
fuzzy measurement. By being {\em close} we shall mean that the 
temporal mean squared deflection  abbreviated according to
\begin{equation}\label{deviation}
\langle (A-a)^2\rangle_T
= \frac{1}{T}\int_0^T\Big( A\big(p(t),q(t),t\big) - a(t) \Big)^2\,dt
\end{equation}
is small.  According to the Feynman approach 
$A\left(p,q,t\right)$ is a $c$-number function corresponding to 
the quantum observable. (\ref{deviation}) is 
an integral characteristic of the difference between two 
functions. This characteristic is large (and the 
curves are therefore not close) only if the deflection of one 
curve from another is big during a long time. The curves which 
have large deflections but only during short intervals are 
considered to be close.\footnote{We accept 
here all curves $[a]$ as possible measurement 
readouts. In fact only comparatively slowly varying curves 
$[a]$ must be considered if 
the inertial properties of a real measuring device are taken into 
account. Then large deflections for a short  time are 
impossible.}

Based on this, the information supplied by the readout $[a]$ of a 
continuous fuzzy measurement can be typically presented by the 
{\em Gaussian weight functional}
\begin{equation}\label{weight}
w_{[a]}[p,q] = \exp\left[ -\kappa \int_0^T
\Big( A\big(p(t),q(t),t\big) - a(t) \Big)^2\,dt \right]
\end{equation}
which is exponentially small if $\langle (A-a)^2\rangle_T$ is 
big. The inverse  of the parameter $\kappa$ may serve as a {\em 
measure of fuzziness}. Later on, it will be connected with the 
resolution of these devices. Eq.~(\ref{weight}) characterizes a 
particular class of measuring devices.\footnote{The case of a 
sharp continuous measurement (when a single path presents 
a measurement readout) is considered in the paper of Aharonov 
and Vardi \cite{Aharon}.}

We shall consider the case when the fuzziness of the measurement 
does not alter with time, so that $\kappa$ is constant. In 
principle, the parameter $\kappa$ could depend on time (then it 
must be included in the integrand of the time integral). In any 
case $\kappa$ cannot depend on the duration $T$ of the 
measurement. This follows from  group-theoretical properties of 
continuous measurements \cite{book93,group-cont-meas}, but can 
also be derived from models of measurements \cite{KMNamiot}. An 
important consequence of this fact will be discussed below,  
namely that the resolution of the continuous measurement is 
improved according to ${{\Delta a}}_T\sim 1/\sqrt{T}$ when its 
duration increases.

\subsection{Complex Hamiltonian}

Using the weight functional (\ref{weight}) in Eq.~(\ref{part-prop}),
we find for the partial propagators describing the
evolution corresponding to a certain measurement output $[a]$
\begin{equation}\label{path-prop}
U_T^{[a]}(q'',q')=\int d[p]\,d[q]\,
\exp\left\{ \frac{i}{\hbar} \int_0^T \big(p\dot q
- H_{[a]}\,(p,q,t)\big)\,dt \right\}
\end{equation}
where $p=p(t)$ and $q=q(t)$. The restricted path integral
(\ref{path-prop}) has the form of a
non-restricted Feynman path integral but with an
{\em effective Hamiltonian}
\begin{equation}\label{effect-Ham}
H_{[a]}\,(p,q,t) = H(p,q,t) - i\kappa\hbar \,\big( A(p,q,t) - a(t)
\big)^2
\end{equation}
containing an imaginary term.\footnote{The form of the imaginary term
may be different for another weight functional $w_{[a]}$, i.e. for
another class of measuring devices.}

The possibility to describe the dynamics by an 
effective Hamiltonian is very important since it makes 
practical applications rather simple: one may 
forget about path integrals and reduce the problem to the 
solution of the equivalent Schr\"odinger equation with a complex 
Hamiltonian. It is important also from the theoretical point of 
view because the imaginary part of the Hamiltonian indicates  a 
characteristic trait of the process of continuous measurement 
namely that   
information is dissipated.\footnote{Let us remark that restricted 
path integrals describing continuous fuzzy measurements and 
leading to non-Hermitian Hamiltonians must not be confused with 
path integration in the presence of infinite potential barriers. 
Potential barriers also do not permit paths to go out of some 
corridor, but the unitary character of the process is not 
violated, no dissipation occurs.}

\subsection{Measurement of energy}

We turn to the measurement of energy. Let the Hamiltonian
 $H$ of the system have the form
$
H=H_0+V
$
where $H_0$ is the Hamiltonian of the ``free" multilevel system 
and $V$ is a potential describing an external influence leading 
to transitions between levels if no measurement is applied. 
By measurement of the energy we 
shall mean the measurement of the observable $A=H_0$ leading to 
the readout $[E]=\{E(t')|0\le t' \le T\}$.

The weight factor (\ref{weight}) with $A=H_0$ and $[a]=[E]$
can be written in the form
\begin{equation}\label{weight-energy}
w_{[E]}[p,q]
= \exp\left[ 
-\frac{\langle (H_0-E)^2\rangle_T}{{{\Delta E}}_T^2}\right]
\end{equation}
where we made use of the notation (\ref{deviation}) and 
introduced the parameter
\begin{equation}\label{DeltaE}
{{\Delta E}}_T = \frac{1}{\sqrt{\kappa T}}
\end{equation}
characterizing the fuzziness of the measurement in a more 
transparent way than $\kappa$. 

The weight factor (\ref{weight-energy}) guarantees that only 
those Feynman paths $[p,q]$ are taken into account, for which the 
time dependence of the observable $H_0$ is presented by a curve 
close to the curve $[E]$. The measure of the deflection $\langle 
(H_0-E)^2\rangle_T$ is, according to (\ref{weight-energy}), less 
or of the order of ${\Delta E}_T^2$. This means that the 
{\em resolution} of the measuring device described by our formulas is 
equal to ${\Delta E}_T$.\footnote{In the preceding papers on RPI 
approach in continuous measurements an analogous parameter was 
usually called measurement error. However, use of this term may 
be misleading, and now we prefer a more definite term 
``resolution".}

It has been stated above that the parameter $\kappa$ does not 
depend on the duration $T$ of the measurement. Therefore, it 
follows from Eq.~(\ref{DeltaE}) that the resolution of the 
continuous fuzzy measurement is improved when the duration of 
measurement increases. This important property may be 
understood if one 
replaces the continuous fuzzy measurement by a sequence of 
instantaneous fuzzy measurements with Gaussian weight 
functionals. It is plausible that such repetition of fuzzy 
measurements improves the resulting sharpness. 

The measurement of energy is now worked out as follows: 
Let the measured system be initially in a pure state 
$|\psi_0\rangle$. Then for a particular given readout $[E]$ its 
further evolution is obtained by solving the corresponding 
Schr\"odinger equation (comp. (\ref{effect-Ham}) with $A=H_0$ 
and $[a]=[E]$) 
\begin{equation}\label{Schroed-select}
\frac{\partial}{\partial t} |\psi_t\rangle
  = \left(-\frac{i}{\hbar} H
  -\kappa \,\Big( H_0 - E(t)\Big) ^2\right)\, |\psi_t\rangle.
\end{equation}
The norm of the final state vector $|\psi_T\rangle$ gives then, 
according to the general formula (\ref{prob}), the probability 
density for the measurement readout $[E]$:
\begin{equation}
P[E]=\langle\psi_T|\psi_T\rangle. 
\end{equation}
This together with the resulting state $|\psi_T\rangle$ is what 
can be known about the measurement. 

Expanding the state $|\psi_t\rangle$ in the basis 
\begin{equation}\label{eigenstates}
|\varphi_n(t)\rangle = e^{-iE_n\, t/\hbar}|n\rangle
\end{equation}
of time-dependent eigenstates of $H_0$, we have the following 
system of equations for the coefficients of the expansion
$|\psi_t\rangle = \sum C_n(t)|\varphi_n(t)\rangle$:
\begin{equation}\label{C-eq-gen}
\dot C_n = -\kappa (E_n-E(t))^2\,C_n
  -\frac{i}{\hbar}
  \sum_{n'}\langle\varphi_n|V|\varphi_{n'}\rangle C_{n'}.
\end{equation}
The probability density of the measurement readouts is 
given by 
\begin{equation}\label{prob-gen}
P[E]=\sum_{n} |C_n(T)|^2. 
\end{equation}

\section{``Free" multilevel system: model of decoherence}\label{free}

We shall consider first a ``free" multilevel system with the 
Hamiltonian $H_0$. ``Free" means vanishing $V$ so that there is no 
driving field and no transitions between levels occur. We
shall see in this simple case that a continuous fuzzy measurement of
energy may serve as a model for decoherence leading to the resolution
of the energy levels.

Because of $V=0$, Eq.~(\ref{C-eq-gen}) have a simple solution
\begin{equation}\label{cn-free}
C_n(T)=C_n(0)\exp \left[ - \kappa\int_0^T dt
\left( E_n - E(t)\right) ^2\right]
=C_n(0)\exp \left[
-\frac{\langle (E_n-E)^2\rangle_T}{{\Delta E}_T^2}\right]
\end{equation}
where $\langle\, \rangle_T$ is a time-average defined in 
(\ref{deviation}) and ${\Delta E}_T$ is defined by 
Eq.~(\ref{DeltaE}). 

\subsection{Probability distribution}\label{estimat}

A probabilistic analysis may be based on 
Eqs.~(\ref{prob}) and (\ref{unitar}). In our case the probability 
density
\begin{equation}
P[E]= \sum_n |C_n(T)|^2
=\sum_n |C_n(0)|^2
\exp\left[ -2\frac{\langle (E_n-E)^2\rangle_T}{{\Delta E}_T^2} \right]
\label{prob-dens}
\end{equation}
is exponentially small when the temporal mean squared deflection
$\langle (E_n-E)^2\rangle_T$ of
the function $[E]$ from each level $E_n$ is more than 
${\Delta E}_T^2$. On the contrary, the probability
density is close to the maximum if $\langle (E_n-E)^2\rangle_T$ is much
smaller than ${\Delta E}_T^2$ for the level $E_n$ corresponding 
to a maximum value of $|C_n(0)|^2$.

Let us now turn to the probability itself. The generalized 
unitarity has in the case of the energy measurement the form
$\int d[E]\, U_{[E]}^{\dag}U_{[E]}= {\bf 1}$. Taking matrix 
elements of this equality in the energy representation, we have 
an equivalent condition
\begin{equation}\label{en-unitar}
\int d[E]\,
\exp \left[ -2\frac{\langle (E-E_n)^2\rangle_T}{{\Delta E}_T^2}\right]
=1
\quad\mbox{(for all $n$)}.
\end{equation}
The path integral here is of Gaussian type and can be evaluated. 
It may be shown that the (generalized) unitarity condition is 
fulfilled if the functional measure $d[E]$ is chosen 
to be 
$
d[E]=\prod_t \sqrt{2\kappa dt/\pi}\,dE(t)
$.

Because of an exponential in the integrand, only paths close 
enough to the level $E_n$ contribute in (\ref{en-unitar}). 
Namely, the integral which is analogous but restricted to the 
subset of readouts 
\begin{equation}\label{An}
{\cal A}_n = 
\{ [E]\, |\, \langle (E-E_n)^2\rangle_T \lesssim {\Delta E}^2_T \},
\end{equation}
is close to unity:
\begin{equation}\label{en-unitar-An}
\int_{{\cal A}_n} d[E]\,
\exp \left[ 
-2\frac{\langle (E-E_n)^2\rangle_T}{{\Delta E}_T^2}\right]
\sim 1.
\end{equation}
The probability that the measurement output $[E]$ belongs to an
arbitrary set ${\cal A}$ is according to the general formula
(\ref{prob}) and with the help of (\ref{prob-dens}) 
\begin{equation}\label{prob-pn}
{\rm {Prob}}([E]\in{\cal A})=
\int_{\cal A} d[E]\,
{\rm Tr}\left( U_{[E]}\,\rho_0 \, U_{[E]}^{\dag}\right)
=\sum_n p_n({\cal A})\, |C_n(0)|^2
\end{equation}
with
\begin{equation}
p_n({\cal A}) =
\int_{\cal A}  d[E] \,
\exp\left[ -2\frac{\langle (E-E_n)^2\rangle_T}{{\Delta E}_T^2} \right].
\end{equation}
Comparing this with Eq.~(\ref{en-unitar-An}), we see that $p_n = 
1$ if ${\cal A}_n\subset{\cal A}$. If, on the contrary, the set 
${\cal A}$ does not intersect with ${\cal A}_n$ (or their 
intersection is small compared with ${\cal A}_n$) then  $p_n=0$.

\subsection{Different Regimes of Measurement}\label{regimes}

As has already been said, the measurement resolution
${\Delta E}_T$ decreases for increasing duration $T$ of the 
measurement. To define different regimes of the measurement,
this resolution should be
compared with the characteristic difference 
$
{\Delta E}\sim E_{n+1}-E_n
$
between energy
levels.\footnote{\label{foot9}If the difference between levels
substantially depends on the number $n$ of the level, one can 
draw corresponding 
conclusions concerning the measurement in a certain energy band
referring to the value ${\Delta E}$ which is typical for this
band.}
It is more convenient however to go over to the corresponding
time parameters.
Let $T_{\rm lr}$ be the duration at which the measurement resolution
${\Delta E}_T$ achieves the value $\Delta E$,
\begin{equation}\label{resolut-time}
T_{\rm lr} = \frac{1}{\kappa{\Delta E}^2}.
\end{equation}
Then ${\Delta E}_T$ is larger or smaller than ${\Delta E}$ depending on
whether $T$ is smaller or larger in comparison with $T_{\rm lr}$.

From Eq.~(\ref{cn-free}) we have for the
coefficients $C_n$ at time $t=T_{\rm lr}$
\begin{equation}\label{cn-free-decoh}
C_n(T_{\rm lr})=C_n(0)\exp \left[
-\frac{\langle (E_n-E)^2\rangle_{T_{\rm lr}}}{{\Delta E}^2}\right].
\end{equation}
Therefore, the coefficient $C_n(t)$ decreases essentially during the
time $T_{\rm lr}$ if the deflection of $E(t)$ from $E_n$ is of the order
or greater than ${\Delta E}$. This is why the parameter $T_{\rm lr}$ may
be called {\em level resolution time}. Additional arguments for this
will be given below.

Let us consider now different durations $T$ and the corresponding 
regimes of measurement.
In the case $T\gg T_{\rm lr}$ we have ${\Delta E}_T\ll{\Delta 
E}$. Now the probability density (\ref{prob-dens}) is 
not negligible only if for most of the measurement duration
\begin{equation}\label{output-probable}
|E-E_n| \,\lesssim\, {\Delta E}_T\ll{\Delta E}
\end{equation}
for some $n$. The corresponding set of readouts ${\cal A}_n$ 
determined by Eq.~(\ref{An}) is in this case a very narrow band  
close to the constant curve $E(t)\equiv E_n$. The width of this 
band is much less than ${\Delta E}$, so that no other level 
$E_{n'}$ with $n'\ne n$ lies inside ${\cal A}_n$. Taking the set of 
paths ${\cal A}$ in (\ref{prob-pn}) in such a way that it 
includes ${\cal A}_n$ but does not intersect with any other 
${\cal A}_{n'}$, $n'\ne n$, we have $p_n=1$ and $p_{n'}=0$ 
for $n'\ne n$. Therefore it follows from (\ref{prob-pn}) that
\begin{equation}
{\rm {Prob}}({\cal A}) = |C_n(0)|^2.
\end{equation}

Thus, in the case $T\gg T_{\rm lr}$ the measurement readout $[E]$ 
will be an almost constant curve close to one of the levels 
$E_n$, and the probability of it being close to a specified $E_n$ 
is $|C_n(0)|^2$. We see that the continuous fuzzy measurement, if 
it lasts longer than $T_{\rm lr}$, distinguishes (resolves) 
between the levels. This justifies once more the name 
{\em level resolution time} for the parameter $T_{\rm lr}$.
To find the state of the system after the measurement in the
considered case ($T\gg T_{\rm lr}$), we have to use Eq.~(\ref{cn-free})
with the
function $E(t)$ satisfying the restriction (\ref{output-probable}). We obtain
\begin{equation}\label{state-Zeno}
C_n(T) = C_n(0),
\quad C_{n'}(T) = 0, \quad n'\ne n.
\end{equation}

Thus, after a continuous fuzzy measurement of duration longer 
than $T_{\rm lr}$ the situation is similar to the one in the 
conventional description of the energy measurement as it is given 
in the von Neumann postulate: 1)~The only possible measurement 
readouts $[E]$ are those close to one of the energy levels $E_n$, 
2)~the probability of $[E]$ being close to a particular $E_n$ is  
$|C_n(0)|^2$ and 3)~in the case of $[E]$ being close to 
$E_n$ the system turns out to be in the energy eigenstate 
$|\varphi_n\rangle$ after the measurement.

The physical process leading to the determination of an energy 
eigenvalue (or of the eigenvalue of any other observable with 
discrete spectrum) has been investigated in the framework of 
different models and was called the process of decoherence 
\cite{JoosZeh,Zurek}. We see now that the continuous fuzzy 
measurement of energy fixes the energy level during $T_{\rm 
lr}$. Therefore it presents a phenomenological 
description of the decoherence process 
leading to the determination of the specific energy eigenvalue. 
The characteristic mark of decoherence is the dying out the 
off-diagonal matrix elements of the density matrix. It is easily 
seen that it is this what really happens at the time $T_{\rm lr}$ 
because due to (\ref{cn-free-decoh}) all products  
$C_iC_j^{*}$ with $i\ne j$ become exponentially 
small for $T\gg T_{\rm lr}$. The time  period during which 
non-diagonal elements die out, is usually called the decoherence 
time. We call it level resolution time because we consider other 
regimes of measurement too, where 
the parameter $T_{\rm lr}$ plays another role. 

In this context the measurement performed during the time $T\ll 
T_{\rm lr}$ might be considered as not yet finished. However this 
point of view is applicable only if the aim of the measurement is 
to distinguish between different levels. Continuous fuzzy 
measurement may however have as a goal to estimate the energy 
of the system with a resolution worse than $\Delta E$. Then a period of 
measurement shorter than $T_{\rm lr}$ makes sense too. Let us 
turn to this regime.

We consider the case $T\ll T_{\rm lr}$. This is equivalent to 
${\Delta E}_T\gg{\Delta E}$, so that the set of paths ${\cal 
A}_n$ determined by Eq.~(\ref{An}) includes functions $[E]$ in a  
very wide energy band around the constant curve $E\equiv 
E_n$. The width of this band is ${\Delta E}_T$, so that it 
includes many other energy levels. Vice versa, if we define 
the set of readouts ${\cal A}$ to contain all functions $[E]$ in 
the energy band of the width ${\Delta E}_T$ between $E_{\rm min}$ 
and $E_{\rm max}$, then many energy levels $E_n$, $n_1\le n\le 
n_2$ lie inside this band. For each of these levels intersection 
between ${\cal A}_n$ and ${\cal A}$ makes a significant part of 
${\cal A}_n$ and therefore $p_n({\cal A}) \sim 1 \mbox{ for } 
n_1\le n\le n_2$. On the other hand, for $n$ outside the interval 
$[n_1, n_2]$ intersection between ${\cal A}_n$ and ${\cal A}$ is 
small so that $p_n({\cal A})$ is small. Making use of 
Eq.~(\ref{prob-pn}), we have approximately
\begin{equation}
{\rm {Prob}}({\cal A}) = \sum_{n=n_1}^{n_2} |C_n(0)|^2.
\end{equation}
This regime of measurement does not distinguish between the 
energy levels so that the character of the measurement in this 
regime is essentially the same as in the case of a continuous 
spectrum.

\section{The driven system: correlation between measurement
readout and Rabi flopping}\label{two-level}

We consider now a system which is exposed to an external 
driving field. For simplicity let it be a 2-level system under 
resonance influence of a periodical force with frequency $\omega 
= \Delta E/\hbar$ where $\Delta E=E_2-E_1$. The potential energy 
has then the following non-zero matrix elements for the states 
(\ref{eigenstates}): $\langle\varphi_1|V|\varphi_2\rangle 
=\langle\varphi_2|V|\varphi_1\rangle^{*}=V_0$. 
Eq.~(\ref{C-eq-gen}) gives for the related expansion coefficients 
\begin{eqnarray} \label{2-level}
\dot C_1 &=&  -i v C_2 - \kappa (E_1-E(t))^2\, C_1,
\nonumber\\
\dot C_2 &=&  -i v C_1 - \kappa (E_2-E(t))^2\, C_2
\end{eqnarray}
with $v={V_0}/{\hbar}$.

If no measurement takes place, i.e. for $\kappa=0$, such a system 
undergoes periodic transitions between the eigenstates ({\em Rabi 
oscillations} or {\em Rabi flopping}):
\begin{equation}\label{Rabi-sol}
C_1(t) =R_1(t), \quad C_2(t) =R_2(t)
\end{equation}
with
\begin{eqnarray}
R_1(t)&=&C_1(0)\,\cos v t -iC_2(0)\,\sin v t, \nonumber\\
R_2(t)&=&C_2(0)\,\cos v t -iC_1(0)\,\sin v t. \label{Rabi-osc}
\end{eqnarray}
The parameter $v=V_0/\hbar$ is equal to a half of the Rabi 
frequency and the corresponding time scale is the {\em Rabi 
period} $T_{\rm R}=\pi/v=\hbar\pi/V_0$. In the time $T_{\rm R}$ 
the system inverts completely from one level to the other and 
returns to the original level.

Now the energy measurement is characterized by three different 
time scales: the level resolution time $T_{\rm lr}$, the Rabi 
period $T_{\rm R}$, and the duration of measurement $T$. 
With regard to physical interpretation one may 
look at them as follows: Small $T_{\rm lr}$ 
represents quick level resolution because of small fuzziness and 
therefore strong Zeno-type influence of the measurement. Small 
$T_{\rm R}$ corresponds to a strong influence of the driving 
field. For large $T$ fuzziness effectively decreases. Let us 
consider different limiting cases for the relations between these 
times which will correspond to characteristic regimes of 
measurement. 

If $T\ll T_{\rm R}$, then the measurement duration is too short 
for the effect of the resonance influence to be observed. In this 
case we have in fact a ``free" system as considered in 
Sect.~\ref{free}. The two possible relations, $T_{\rm lr}\ll T$ 
and $T\ll T_{\rm lr}$, between the remaining parameters 
correspond to the cases when the measurement distinguishes 
between the levels and when it is too short to distinguish 
between themes discussed above in Sect.~\ref{free}.

We turn now to the opposite inequality, $T_{\rm R}\lesssim T$. 
Now the influence of the driving field becomes stronger. But 
how big the resulting influence on the state of the system is, 
still depends on the amount of fuzziness of the measurement. 
We will discuss in detail the regimes $T_{\rm lr}\ll T_{\rm R}$ 
(Zeno regime in which influence of the measurement is strong 
enough to damp out the Rabi oscillations), the opposite regime 
$T_{\rm R}\ll T_{\rm lr}$ (Rabi regime in which Rabi oscillations 
maintain but there is no correlation between them and the 
measurement readout $[E]$) and the regime 
$T_{\rm R} < 2\pi T_{\rm lr}$ but with $T_{\rm R}$, 
$2\pi T_{\rm lr}$ and $T$ of the same order. The latter is 
according to our goal the most interesting one, because the 
energy measurement readout  turns out to be correlated to 
the oscillations of the state vector (modified Rabi oscillations). 
Some remarks concerning the regime $T_{\rm R}\ll T_{\rm lr} \ll T$ 
will be made. 

\subsection{Zeno regime of measurement}\label{sec-Zeno}

The first regime of measurement corresponds to the relation 
$T_{\rm lr}\ll T_{\rm R}\ll T$. One of its specific features 
results from the inequality $T_{\rm lr}\ll T$ equivalent to 
${\Delta E}_T\ll {\Delta E}$ or $\kappa T{\Delta E}^2\gg 1$. 
Because of this relation the damping terms in Eq.~(\ref{2-level}) 
act effectively when $[E]$ differs from the corresponding level 
$E_1$ or $E_2$ by a value of the order of ${\Delta E}$ or larger. 
This leads to a decay of the coefficients $C_1$, $C_2$. According 
to (\ref{prob-gen}) the corresponding measurement readout $[E]$ 
has low probability density.

This decay of the coefficients may be prevented (and therefore 
the output may have large probability density) only if $[E]$ is 
close to one of the levels $E_1$ or $E_2$. Recall our definition 
of being close. Intervals when $E(t)$ deflects from the level 
more  than by ${\Delta E}$ may occur because of the fuzziness of 
the measurement, but their complete duration should be much less 
than $T_{\rm lr}$ and therefore much less than $T$.

To verify that the probability density is actually large for and 
$[E]$ close to $E_n$, we consider a solution of 
Eq.~(\ref{2-level}) for the case $E(t)\equiv E_1$.\footnote{We 
simplify the consideration in this way, but the results are 
correct for the function $[E]$ close, not identical  to level 
$E_1$.} Then Eq.~(\ref{2-level}) reads:
\begin{equation}
\dot C_1 =  -i \frac{\pi}{T_{\rm R}} C_2, \quad
\dot C_2 =  -i \frac{\pi}{T_{\rm R}} C_1 - \frac{1}{T_{\rm lr}}\, C_2.
\label{discrete-eq}
\end{equation}
These equations are easily solved  to give
\begin{equation}
C_1=a\, e^{\lambda_1 t}+b\, e^{\lambda_2 t}, \quad
C_2=i\frac{T_{\rm R}}{\pi}
\left( \lambda_1 a\, e^{\lambda_1 t}
+ \lambda_2 b\, e^{\lambda_2 t} \right)
\label{discrete-sol}
\end{equation}
where
\begin{equation}
\lambda_{1,2}=-\frac{1}{2T_{\rm lr}}
   \pm\sqrt{\frac{1}{4T_{\rm lr}^2}-\frac{\pi^2}{T_{\rm R}^2}}
\end{equation}
and the coefficients $a$, $b$ are determined by the initial 
conditions. For $T_{\rm lr}<T_{\rm R}/2\pi$ we have exponentially 
decaying solutions. They can be completely described in a generic 
case, but we shall consider only the limiting regime $T_{\rm 
lr}\ll T_{\rm R}$ when the level resolution is much more rapid 
than the Rabi oscillation. We have in this case
$\lambda_1=-1/T_{\rm lr}$, 
$\lambda_2=-\nu^2\lambda_1$,
where $\nu=\pi T_{\rm lr}/T_{\rm R}$ is a small parameter, and 
the coefficients take the form
\begin{equation}
C_1(T)=(C_1(0)-i\nu C_2(0))\,e^{-\nu^2\frac{T}{T_{\rm lr}}},\quad
C_2(T)=-i\nu C_1(T).
\end{equation}

Estimating the probability density by the formula
(\ref{prob-gen}), we find
\begin{equation}
P(E\equiv E_1)
=(|C_1(0)|^2+\nu^2|C_2(0)|^2)\,e^{-2\nu^2\frac{T}{T_{\rm lr}}}.
\end{equation}
The exponential factor here is close to unity for a very wide 
range of $T$, up to the values of the order of $T_{\rm 
R}^2/T_{\rm lr}$. Omitting this factor and neglecting the term 
containing $\nu^2$, we have
\begin{equation}
P[E\equiv E_1]=|C_1(0)|^2.
\end{equation}
In the same approximation we have 
\begin{equation}
C_1(T)=C_1(0), \quad C_2(T)=0.
\end{equation}

We calculated the probability density for the constant curve $(E\equiv 
E_1)$ being the measurement readout. The same conclusion will be 
valid for the curves close but not identical to this constant 
curve. The set of such curves form a narrow band of the width 
${\Delta E}_T$. Because of ${\Delta E}_T\ll {\Delta E}$ this band 
does not include the second level $E_2$. Just as in 
Sect.~\ref{regimes}, the probability of the measurement output to 
belong to this band is equal to $|C_1(0)|^2$. It is clear that an 
analogous consideration may be applied for the measurement 
readout close to the energy level $E_2$.

Thus, for $T_{\rm lr}\ll T_{\rm R}\ll T$ we have found a very 
simple picture:
i)~Only those measurement outputs $[E]$ have high probability which
are close to one of the constant curves $E\equiv E_1$ and 
$E\equiv E_2$;
ii)~the probability of the output to be close to $E_1$ or $E_2$ is
given by the initial values of the decomposition coefficients 
$|C_1(0)|^2$ or $|C_2(0)|^2$ correspondingly;
iii)~in case of the output being close to $E_1$ or $E_2$ the
final state is correspondingly the eigenstate $|\varphi_1\rangle$ 
or $|\varphi_2\rangle$;
iv)~Rabi oscillations are completely damped away.
It is evident that this picture reflects the dominant influence 
of the quantum Zeno effect. It is therefore justified 
to call this regime the {\em  Zeno regime}.

\subsection{Rabi regime of measurement}\label{Rabi}

We turn now to the case $T_{\rm R}\lesssim T_{\rm lr}$ when Rabi
oscillations are not damped. The three variants of this regime 
will be discussed. 

1. We consider first the case $T_{\rm R}\ll T\ll T_{\rm lr}$. 
What can be said about the measurement output $[E]$? Returning to 
Eq.~(\ref{2-level}), we have to apply now the inequality $\kappa 
T{\Delta E}^2\ll 1$ equivalent to $T\ll T_{\rm lr}$. It shows 
that the damping terms are essential only for very large 
deviation of $[E]$ from both $E_1$ and $E_2$. If $[E]$ lies in 
the band of the width ${\Delta E}_T={\Delta E}\sqrt{T_{\rm 
lr}/T}$ around $E_1$, $E_2$, no damping occurs. The solution of 
Eqs.~(\ref{2-level}) have then the form of the Rabi oscillations 
(\ref{Rabi-sol}) and the probability density for each of these 
measurement outputs $[E]$ is equal (up to the order of magnitude) 
to unity:
\begin{equation}
{\rm Prob}[E] = |R_1(T)|^2 + |R_2(T)|^2 = 1.
\end{equation}
All the curves $[E]$ in the mentioned wide band may 
(approximately) be considered as equally probable.  Probability 
density essentially decreases only for the curves which deviate 
from both levels $E_1$, $E_2$ more than by the amount ${\Delta 
E}_T$ (and therefore much more than by ${\Delta E}$). The 
measurement in this regime does not respect the discrete 
character of the energy spectrum. The duration of the measurement 
is too short to resolve the levels. The energy is measured with 
the error ${\Delta E}_T$ much greater than ${\Delta E}$.

Turning to the corresponding state of the system we note that 
with the damping terms being negligible in the present regime, 
Eqs.~(\ref{2-level}) have a very simple solution (\ref{Rabi-sol}) 
describing Rabi oscillations. These oscillations are not 
prevented by the continuous measurement performed in this regime. 
They are sinusoidal with Rabi period $T_{\rm R}$. We shall call
therefore this regime of measurement the {\em  strong Rabi regime}. 
The oscillations of the atom state are not reflected in this case in the 
readout $[E]$. 

2. The important variant of the Rabi regime is the one which 
shows a correlation between oscillations of the state vector 
and the measurement readout. We shall demonstrate that this  
takes place for $T_{\rm R} < 2\pi T_{\rm lr}$ but with $T_{\rm R}$, 
$2\pi T_{\rm lr}$ and $T$ being of the same order. . The analysis of 
Eqs.~(\ref{2-level}) is more complicated and less reliable in 
this case, but the process may be investigated with the help of 
the numerical simulation.

Let us rewrite Eq.~(\ref{2-level}) in the form
\begin{equation}\label{2-lev-2}
\dot C_1 =  -i v C_2 - {\epsilon}_1\, C_1,
\quad
\dot C_2 =  -i v C_1 - {\epsilon}_2\, C_2
\end{equation}
with
\begin{equation}
{\epsilon}_i \left(t \right)= \kappa (E_i-E(t))^2
=\frac{1}{T_{\rm lr}}\left( \frac{E-E_i}{{\Delta E}}\right)^2.
\end{equation}
Introduce instead of $C_n$ the new functions:
\begin{equation}
S_n(t) =e^{{\cal E}(t)}C_n(t)
\end{equation}
with
\begin{equation}
{\cal E}(t)=\frac 12 \int_0^t ({\epsilon}_1(t)+{\epsilon}_2(t))\, dt.
\end{equation}
Then these functions satisfy the equations
\begin{equation}\label{2-lev-S}
\dot S_1 +i  v S_2 + \frac 12 {\sigma} S_1 = 0,
\quad
\dot S_2 +i  v S_1 - \frac 12 {\sigma} S_2 = 0
\end{equation}
whereby ${\sigma} = {\epsilon}_1 - {\epsilon}_2$. 
Eqs.~(\ref{2-lev-S}) imply the decoupled second-order equations: 
\begin{eqnarray}\label{S-2ord}
\ddot S_1+(v^2+\frac 12 \dot{\sigma}-\frac 14 {\sigma}^2)\,S_1
&=& 0, \nonumber\\
\ddot S_2+(v^2-\frac 12 \dot{\sigma}-\frac 14 {\sigma}^2)\,S_2
&=& 0.
\end{eqnarray}

We shall assume (this will be justified {\em a posteriori} for 
the most probable measurement readout $[E_{\rm prob}]$) that 
i)~the deviation $|E(t)-\bar E |$ with $\bar E=\frac 12 (E_1+E_2)$ 
is not much larger than 
$\Delta E$ during most of the time, i.e. goes beyond these limits 
only for a short time or as a small deviation and ii)~the 
spectrum of the function $E(t)$ does not contain frequencies much 
higher than $v$ (a half of the Rabi frequency). Then
\begin{equation}
\frac{|{\sigma}|}{v}\lesssim \frac{T_{\rm R}}{2\pi T_{\rm lr}},
\quad
\frac{|\dot{\sigma}|}{v^2}\lesssim \frac{T_{\rm R}}{2\pi T_{\rm lr}}.
\end{equation}
Accordingly, to zeroth order in the ratio $T_{\rm R}/2\pi T_{\rm 
lr}$ we neglect the two last terms in the brackets of 
Eq.~(\ref{S-2ord}). The functions $S_n$ are  then approximated 
by harmonic functions of frequency $v$, i.e. by the Rabi 
oscillations $R_n$ of (\ref{Rabi-osc}).

We have therefore as a zeroth approximation
\begin{equation}\label{0-approx}
C_1(t)=e^{-{\cal E}(t)}R_1(t), \quad
C_2(t)=e^{-{\cal E}(t)}R_2(t).
\end{equation}
This approximation reflects the general oscillatory character of 
the evolution of the state $|\psi_t\rangle$, but it is too rough 
to use it for calculating the probability and for estimating the most 
probable measurement readout $[E]$. Instead, we shall use another 
procedure for this.

Consider again the functions $C_n(t)$ at an arbitrary time moment
$0\le t \le T$ and form the function
\begin{equation}
P(t)=|C_1(t)|^2 + |C_2(t)|^2.
\label{prob-t}\end{equation}
The quantity $P(t)$ depends on the values $\{ E(t')|0\le t'\le 
t\}$ of the function $[E]$ in the moments preceding $t$. This 
quantity can be interpreted as the probability density of the 
corresponding measurement readout if the measurement is finished 
at time $t$. With increasing time $t$ the value $P(t)$ 
decreases.

Let us analyze how the function $P(t)$ depends on time and choose 
the function $[E]$ in such a way that $P(t)$ decreases as slowly 
as possible. Then the resulting function $[E]$ will correspond to 
the most probable measurement readout.

Differentiating Eq.~(\ref{prob-t}) and making use of the
Eq.~(\ref{2-level}) we have
\begin{equation}
\dot P(t) = -2\kappa \left[(E_1-E(t))^2\,|C_1|^2
+ (E_2-E(t))^2\,|C_2|^2\right].
\label{P-dot}\end{equation}
We minimize the absolute value of the r.h.s. and obtain
for the most probable measurement output
\begin{equation}
E_{\rm prob}(t)
= \frac{E_1|C_1(t)|^2 + E_2|C_2(t)|^2}{|C_1(t)|^2 + |C_2(t)|^2}.
\label{E-prob}\end{equation}

Now we can replace $C_n$ by the approximate solution 
(\ref{0-approx}). This gives for the most probable measurement 
output
\begin{equation}
E_{\rm prob} (t)
= E_1|R_1(t)|^2 + E_2|R_2 (t)|^2.
\label{E-prob-Rabi}\end{equation}
This curve is correlated with the Rabi oscillations of the state 
in the sense that the curve $E_{\rm prob}(t)$ is closer to the 
level $E_n$ when the probability for the system to be on this 
level is larger. The function $E_{\rm prob}(t)$ oscillates around 
the middle line $\bar E$ between the levels. The amplitude of the 
oscillations is equal to $\frac 12{\Delta E} \sqrt{(|C_1(0)|^2 - 
|C_2(0)|^2)^2 +4({\rm Im}(C_1 C_2^{*}))^2}$. If the initial 
value of the decomposition coefficient 
$C_1(0)$ is equal to $C_2(0)$, the most probable measurement 
output $E_{\rm prob}$ coincides with the constant curve 
$E\equiv\bar E$. Only in this case the Rabi oscillations 
disappear.

The measurement readout (\ref{E-prob-Rabi}) is shown to be most 
probable. However, it is not clear from the above argument how 
rapidly the probability density decreases if the readout $[E]$ 
deviates from its most probable shape. If it decreases rapidly 
enough, then we may speak about correlation between the 
state oscillations and measurement readout. This is 
investigated with the help of the numerical calculation. We omit 
details. The following conclusions are obtained: 
\begin{description}

\item[i)] Oscillations of the state vector between the energy eigenstates  
(Rabi regime) takes place for $T_{\rm R}\lesssim 2\pi T_{\rm lr}$. 
The boundary to the Zeno 
regime is marked by $T_{\rm R}\approx 2\pi T_{\rm lr}$. 

\item[ii)] The state oscillations are modified Rabi oscillations.  
Their frequency becomes smaller than the Rabi frequency when  
$T_{\rm R}$ approaches  $2\pi T_{\rm lr}$. Together with this 
the pure sinusoidal shape of the oscillations is lost. This indicates 
that there is already some Zeno-type influence of the 
measurement. See the figure for more details. 

\item[iii)] Correlation between the modified Rabi oscillations and the 
measurement readout $[E]$ is substantial for $T_{\rm R}$ close to $2\pi 
T_{\rm lr}$. In the other case the band of different readouts $[E]$ 
having large probability becomes too wide to speak about a 
correlation.

\end{description}

3. The complex case $T_{\rm R}\ll T_{\rm lr}\ll T$ is not treated
in detail here. We restrict ourselves to some comments. 
In this case small errors related to the approximate character of 
(\ref{0-approx}) grow exponentially, so that this solution is 
inappropriate for the analysis, and the conclusions made above 
are not valid. The evolution is qualitatively different in this 
case. Under these conditions the system ``forgets" because of 
the measurement its initial conditions and the oscillating curve 
(\ref{E-prob-Rabi}) cannot arise as a typical measurement output.

The mechanism of ``forgetting" is loss of coherence leading to 
a mixture instead of a pure state. If the duration $T$ of the 
measurement is not too long, the result of the measurement and 
the evolution of the system may be adequately presented by a 
single readout $[E]$ and the resulting pure final state 
$|\psi^{[E]}_T\rangle$. A single readout is a set of null measure 
and strictly speaking is characterized by zero probability. One 
has to consider a band of readouts instead and integrate the 
probability density $P[E]$ over this band to obtain non-zero 
probability. However close measurement readouts $[E]$ lead to 
close final states of the system if the measurement is not too 
long. Therefore, the final state corresponding to the band of 
readouts will be the same pure state  $|\psi^{[E]}_T\rangle$ 
which corresponds to each of the readouts in the band. 

This argument is not valid however for a long measurement. In 
this case even very close readouts $[E]$ lead to different final 
states $|\psi^{[E]}_T\rangle$. Therefore, even a very thin band 
of readouts gives a mixed final state. With increasing $T$ this 
mixed state becomes universal, not depending on the initial state 
of the system.

Indirectly the latter statement is confirmed by the non-selective 
description of the measurement with the help of the density 
matrix (\ref{nonselect-evol}). It has been shown in 
\cite{Master-eq} that this density matrix satisfies a master 
equation having the form of the Liouville equation but with an 
additional term proportional to the double commutator 
$[A[A\rho]]$. In our case this equation reads as follows
\begin{equation}
\dot{\rho} = -\frac{i}{\hbar} [H_0+V,\rho]
-\frac{\kappa}{2}[H_0[H_0\rho]]
\end{equation}
and can easily be solved for a resonantly driven two-level system.
This gives for $8\pi T_{\rm lr}>T_{\rm R}$ (Rabi regime)
\begin{equation}
\rho_{11}-\rho_{22}
= e^{- (t/4T_{\rm lr})} (a\cos \Omega t + b \sin \Omega t).
\end{equation}
Here matrix elements are taken between the states 
(\ref{eigenstates}), $\Omega=\sqrt{(2v)^2-(2T_{\rm lr})^{-2}}$ 
and the parameters $a$, $b$ are determined by initial conditions. 
It is evident that the difference $\rho_{11}-\rho_{22}$ due to 
non-symmetrical initial conditions disappears during time of the 
order of $T_{\rm lr}$.

\section{Possible realizations of continuous fuzzy measurements of
energy}\label{realize}

We want to sketch briefly possible realizations of continuous 
fuzzy measurements of energy. Our suggestions need more 
detailed elaboration which cannot be given here. 
An output of an instantaneous 
measurement is a number. An output of a sequential measurement is 
a series of numbers $a_1, a_2, \dots , a_n, \dots$. If the time scale
is larger than the interval between instantaneous measurements this 
series of numbers may be identified with a continuous curve $a(t)$ 
presenting the readout of a continuous measurement (monitoring).  
Fuzziness of the continuous measurement depends on
 the fuzziness of the instantaneous measurements and on the 
intervals between them. We give three examples: 

1. The most obvious approach is to make a continuous sharp 
measurement fuzzy in making the underlying instantaneous 
measurements fuzzy, in other words, to modify measurements 
leading to the quantum Zeno effect in this way. 
An arrangement for the Zeno effect was 
suggested in \cite{Cook} and used in \cite{Itano}. A system (ion 
in a trap) consisting of levels 1 and 2 is driven between these 
levels by a resonant perturbation. There is a subsidiary level 3 
that can decay only to level 1. The state measurement is carried 
out by action of an optical pulse in resonance between 
levels 1 and 3. This pulse can induce the $1\rightarrow3$ 
transition with the subsequent spontaneous return to level 1 
accompanied by the emission of a photon with the same frequency. 
It causes a projection of the ion onto level 1 accompanied by 
scattering the photon. This happens with the probability 
$|c_1|^2$ if the initial state of the ion is 
$c_1|1\rangle + c_2|2\rangle$. With the probability $|c_2|^2$ 
the pulse projects the ion onto level 2 with no 
scattering. 

The measurement of this type 
is sharp (of von Neumann type) and answers the question 
whether the ion is on level 1. Frequent repetition of this 
measurement was shown \cite{Itano} to lead to the quantum 
Zeno effect  freezing the atom on level 1. Each measurement 
in a series completely resolves in this case between the 
energy levels, i.e. is not fuzzy. It is necessary to modify the 
experimental setup to realize a fuzzy measurement of 
energy.\footnote{The repeated measurement of the same 
type but with varying parameters of the system has been 
considered in \cite{MilburnGa}. It is however difficult 
to compare the results obtained in this paper with our results 
because fuzziness was not the aim of investigation in 
\cite{MilburnGa}.}

To do this, one may choose an ion with very fast decay 
of the subsidiary level 3 to both main levels 1 and 2 and 
a continuous spectrum of the radiation inducing the 
transitions $1\rightarrow3$ and $2\rightarrow3$. The width of 
level 3 and the width of the spectrum of the inducing radiation 
have to be more than the difference $\Delta E$ between levels 1 
and 2. The induced transition from levels 1 and 2 to the 
subsidiary level 3 with the subsequent spontaneous transition 
back to the same level will be accompanied by elastic scattering 
of a photon. Measuring the frequency of the scattered photon is 
equivalent in these conditions to an instantaneous fuzzy 
measurement of energy (performed with the resolution worse than 
$\Delta E$). A series of scatterings is then a continuous fuzzy 
measurement of energy. Resolution of the continuous measurement 
$\Delta E_T$ depends on the length of the series or duration $T$ 
of the continuous measurement.
Special precaution should be taken in this scheme to provide that 
only transitions $1\rightarrow3\rightarrow1$ and 
$2\rightarrow3\rightarrow2$ might occur. For this, level 3 
may be double, consisting of two very close levels $1'$ and $2'$ 
with the only transitions $1\leftrightarrow1'$ and 
$2\leftrightarrow2'$ permitted. 

2. An electron in a magnetic field is a two-level system and 
repeated measurements of its energy can be performed with the help of 
several Stern-Gerlach (SG) devices.  
The position of the electron after scattering in the SG device depends  
on its energy. Therefore measuring the final position 
of the electron is in fact measuring  its energy. If the initial position 
is uncertain to some degree, the resulting measurement of energy 
is fuzzy. In fixing the initial uncertainty, we can determine the fuzziness 
of the instantaneous measurement of energy.  If now the electron 
is scattered subsequently in a series of SG devices, we have in fact 
a series of fuzzy instantaneous measurements of energy approximating 
a continuous one. 

As a driving field we can 
include an external electromagnetic field in resonance with the 
difference between the electron levels. This induces periodic transitions 
between the levels. Then all the effects discussed in Sect.~\ref{two-level} 
can be observed with such a system. An analysis is given in 
\cite{Peres-SG,Peres-book}. The results are in agreement with our 
general results. 

3. One more way of introducing fuzziness can be found 
in quantum optics. The multilevel ``free" system is in this case 
realized as a mode of the electromagnetic field. The measurement of 
energy of this system (usually referred to as the measurement of the 
photon number) is realized by the interaction of this mode with 
another (probe) electromagnetic wave through a nonlinear optical 
Kerr medium in which both waves propagate \cite{Yamamoto}. The 
interaction Hamiltonian has the form
\begin{equation}
H_I=\hbar\chi \,(a^{\dagger}a)\,(b^{\dagger}b)
\end{equation}
where $a$, $b$ are the annihilation operators of the two modes. The 
observable to be measured indirectly is $H_0=a^{\dagger}a$, but 
the directly measured observable is a quadrature phase component
\begin{equation}
b_2=\frac{1}{2i}(b-b^{\dagger}).
\end{equation}

After the probe wave has travelled the 
distance $L$ in the Kerr medium, the value of this component is, 
due to the interaction $H_I$, 
\begin{equation}
b_2(L)\simeq \frac{\chi L}{c}\langle b_1(0)\rangle H_0
+ \Delta b_2(0)
\end{equation}
where $\Delta b_2(0)$ is an initial uncertainty of the amplitude 
$b_2$ and $\langle b_1(0)\rangle$ an initial mean value of the 
second quadrature component $b_1=\frac{1}{2}(b+b^{\dagger})$. 
This means that the output for the measurement of $H_0$ is read off 
as the value of the component $b_2$. 
Fuzziness of the measurement depends therefore on $\Delta b_2(0)$ 
and can be controlled. The series of the measurements of this 
type may be considered as a continuous fuzzy measurement of 
energy $H_0$.

\section{Conclusions}\label{conclus}

Fuzziness of a measurement is caused by the low resolution 
of the detector. Such a measurement is unable to transfer the 
state of the system 
to an eigenstate of the measured observable, and the 
measurement readout must not agree with an eigenvalue. 
Such a continuous fuzzy measurement as monitoring an 
observable may be approximated by as a series of unsharp or 
imperfect ``instantaneous" measurements. It is qualitatively 
different from a series of sharp (von Neumann type) measurements. 

We introduced fuzziness with the help of restricted path integrals, 
on the basis of the resulting non-Hermitian  Hamiltonians. We 
discussed the continuous fuzzy measurement (monitoring) of 
energy for a multilevel and in more detail for a two-level system 
leading to a readout $E(t)$. In addition to the duration $T$ of the 
measurement there are two other characteristic time scales.  
The level resolution time $T_{\rm lr}$ represents the strength of 
an inherent Zeno-type influence (``level freezing"). If an external 
resonant driving field is acting upon the system, the strength of 
its influence is represented by the Rabi period $T_{\rm R}$. 
According to the relative orders of magnitudes of these times, 
different regimes with characteristic properties of the measurement
 readout and the related evolution of the state vector can be 
distinguished. 

We obtained the following results: 
\begin{description}

\item[i)] The study of a ``free" multilevel system which is 
only under the influence of the measurement (no driving field 
present) shows that continuous fuzzy measurement of energy 
is a model for decoherence leading to the resolution of separate 
levels.

\item[ii)] The discussion of a resonantly driven two-level system 
reveals two different regimes of measurement: In the Zeno regime 
the oscillations of the state vector are prevented and the 
measurement readouts are essentially discrete. In the Rabi 
regime the state oscillations maintain and the readouts may 
be arbitrary curves. 

\item[iii)] An important new effect is predicted: For time scales 
corresponding to the Rabi regime but being close to the border with 
the Zeno regime the frequency of the state oscillations is 
decreased and their shape deviates from pure Rabi oscillations. 
Correlation are found between the oscillations of the state 
vector and the measurement readouts. Thus, in following 
$E(t)$ one can directly read off the oscillations of the state 
vector. 

\end{description}

Possible ways of realization of a fuzzy continuous measurement of 
energy are shortly commented. 

\vspace{0.5cm}
\centerline{\bf ACKNOWLEDGEMENT}

The authors thank P.Marzlin and V.Namiot for discussions 
regarding the realization of the measurement and F.Schreck for 
the help in numerical simulations. This work was supported in 
part by the Deutsche Forschungsgemeinschaft.

\newpage

\centerline{\large\bf Figure capture}

Correlation between modified Rabi oscillations of the state 
of a resonantly driven two-level system and the readouts of a continuous fuzzy 
measurement of energy. We assume $T_{\rm R}/2\pi T_{\rm lr}=0.8$. 
a) The in the upper diagram shows the periodic oscillations 
of the state vector between the two energy eigenstates. The period is 
close to but larger than the Rabi period $T_{\rm R}$. The shape is of a 
slightly modified Rabi oscillation. b) The diagram in the middle 
shows as the dashed line the most probable energy readout 
$[E_{\rm prob}]$ which is strongly correlated to the state oscillation.  
Two bands around it (with the width $W=2$ and $W=3$ are drawn. 
c)The probability $P$ that an energy readout $[E]$ lies in the band 
with the width $W$ around the most probable readout is given in the 
lower diagram. 


\begin{thebibliography}{10}


\bibitem{Zeno} B. Misra and E. C. G. Sudarshan, J.Math.Phys. {\bf 
18}, 756 (1977); C. B. Chiu, E. C. G. Sudarshan and B. Misra, 
Phys. Rev. {\bf D 16}, 520 (1977); A. Peres, Amer. J. Phys. {\bf 
48}, 931 (1980).

\bibitem{Itano} W. M. Itano, D. J. Heinzen, J. J. Bollinger and 
D.J. Wineland, Phys. Rev. {\bf A 41}, 2295 (1990).


\bibitem{Peres-book} A. Peres, {\em Quantum Theory: Concepts and 
Methods}, Kluwer Academic Publishers, Dordrecht, Boston \& 
London, 1993.


\bibitem{Zeh} H. D. Zeh, in: {\em On the irreversibility of time and 
observation in quantum theory}, Enrico Fermi School of Physics 
{\bf IL}, Academic Press, 1971, p.263.

\bibitem{cont-meas-count} E. B. Davies, {\em Quantum Theory of 
Open Systems}, Academic Press: London, New York, San Francisco, 
1976; M. D. Srinivas, {\em J.Math. Phys.} {\bf 18}, 2138 (1977).

\bibitem{M79} M. B. Mensky, Phys. Rev. {\bf D 20}, 384 (1979); 
Sov. Phys.-JETP {\bf 50}, 667 (1979).

\bibitem{Barch} A. Barchielli, L. Lanz and G. M. Prosperi, Nuovo 
Cim., {\bf B 72}, 79 (1982).

\bibitem{Peres-SG} A. Peres, Continuous monitoring of quantum 
systems, in {\em Information Complexity and Control in Quantum 
Physics}, ed. by A. Blaquiere, S. Diner, and G. Lochak, Springer 
Verlag, Wien, 1987, pp. 235-240.

\bibitem{Milburn-en} D. F. Walls and G. J. Milburn, Phys. Rev. 
{\bf A 31}, 2403 (1985); G. J. Milburn, J. Opt.Soc. Am. {\bf B 
5}, 1317 (1988).

\bibitem{MilburnGa} M. J. Gagen and G. J. Milburn, Phys. Rev. 
{\bf A 47}, 1467 (1993).

\bibitem{Milburn} D. F. Walls, M.J. Collett, and G. J. Milburn, 
Phys. Rev. {\bf D 32}, 3208 (1985); G. J. Milburn, Phys. Rev. 
{\bf A 36}, 5271 (1987); H. M. Wiseman and G. J. Milburn, Phys. 
Rev. {\bf A 47}, 642 (1993); M. J. Gagen, H. M. Wiseman, and G. 
J. Milburn, Phys. Rev. {\bf A 48}, 132 (1993).

\bibitem{JoosZeh} E. Joos, and H. D. Zeh, Z.Phys. {\bf B 59}, 223 
(1985).

\bibitem{Diosi} L. Diosi, Phys.Lett. {\bf A 129}, 419 (1988).

\bibitem{KMNamiot}  A. Konetchnyi, M. B. Mensky and V. Namiot, 
Phys. Lett. {\bf A 177}, 283 (1993).


\bibitem{book93} M. B. Mensky, {\em Continuous Quantum 
Measurements and Path Integrals}, IOP Publishing: Bristol and 
Philadelphia, 1993.


\bibitem{OnoPre93} R. Onofrio, C. Presilla, and U. Tambini, Phys. 
Lett. {\bf A 183}, 135 (1993).

\bibitem{OnoPre95} U. Tambini, C. Presilla, R. Onofrio, Phys. 
Rev. {\bf A 51}, 967 (1995).


\bibitem{Feynman48} R. P. Feynman, Rev. Mod. Phys. {\bf 20}, 367 
(1948).


\bibitem{Aharon}Y. Aharonov and M. Vardi, Phys. Rev. {\bf D21}, 
2235 (1980).


\bibitem{group-cont-meas} M. B. Mensky, Phys. Lett. {\bf A 150}, 
331 (1990).


\bibitem{Zurek} W. H. Zurek, Phys. Rev. {\bf D 24}, 1516 (1981); 
 {\bf D~26}, 1862 (1982).


\bibitem{Master-eq} M.B.Mensky, Phys. Lett. {\bf A~196}, 159 
(1994).


\bibitem{Cook} R.J. Cook, Phys. Scr. {\bf T~21}, 49 (1988). 

\bibitem{Yamamoto} Y.Yamamoto, The Transactions of the IEICE, 
 {\bf E~73}, 1598 (1990).

\end{thebibliography}
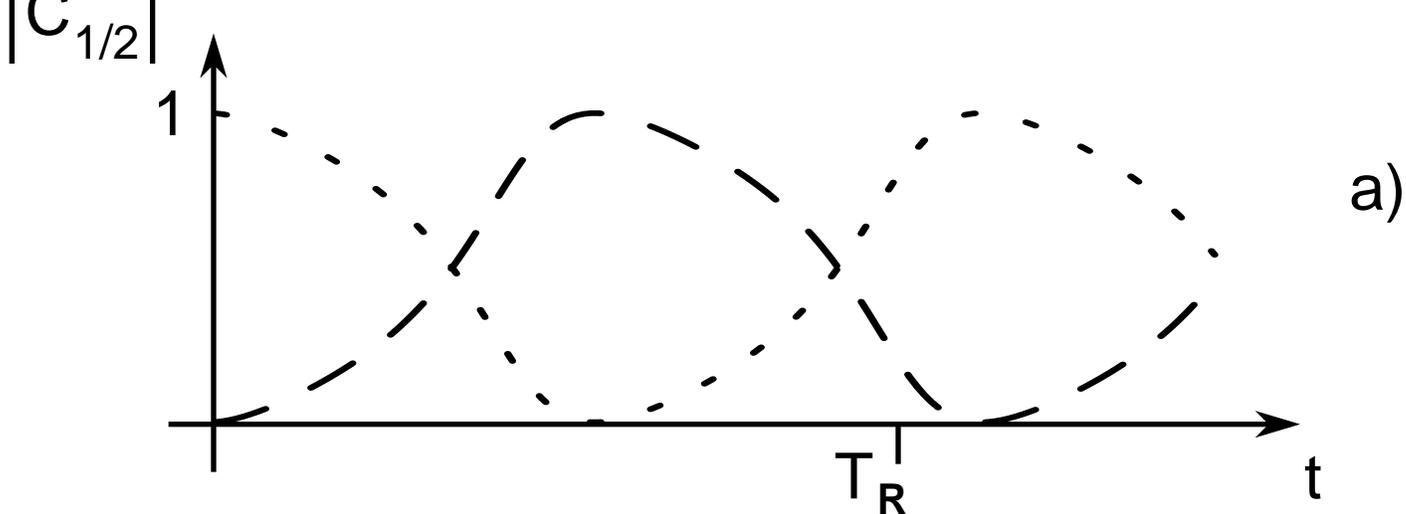
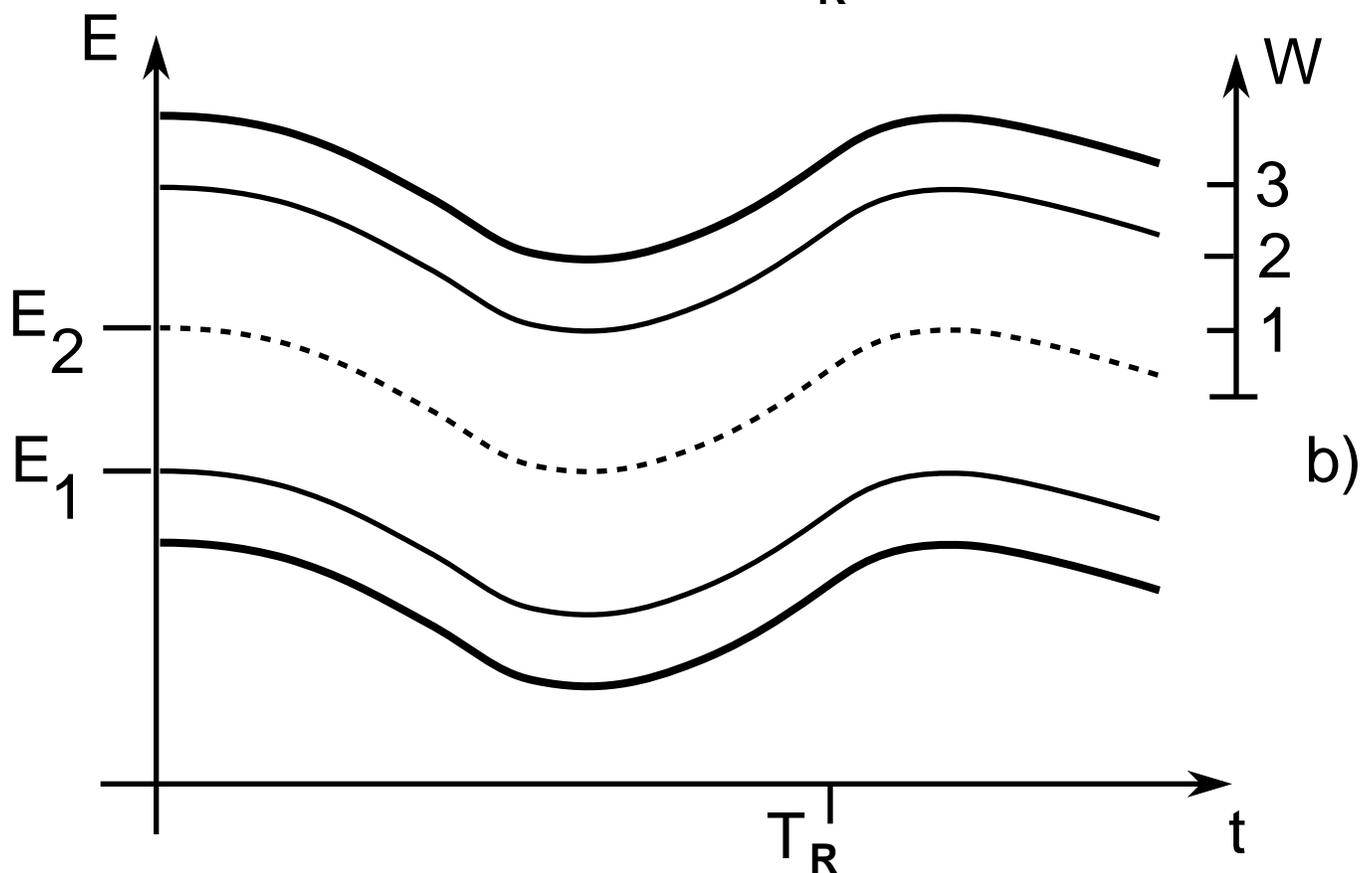
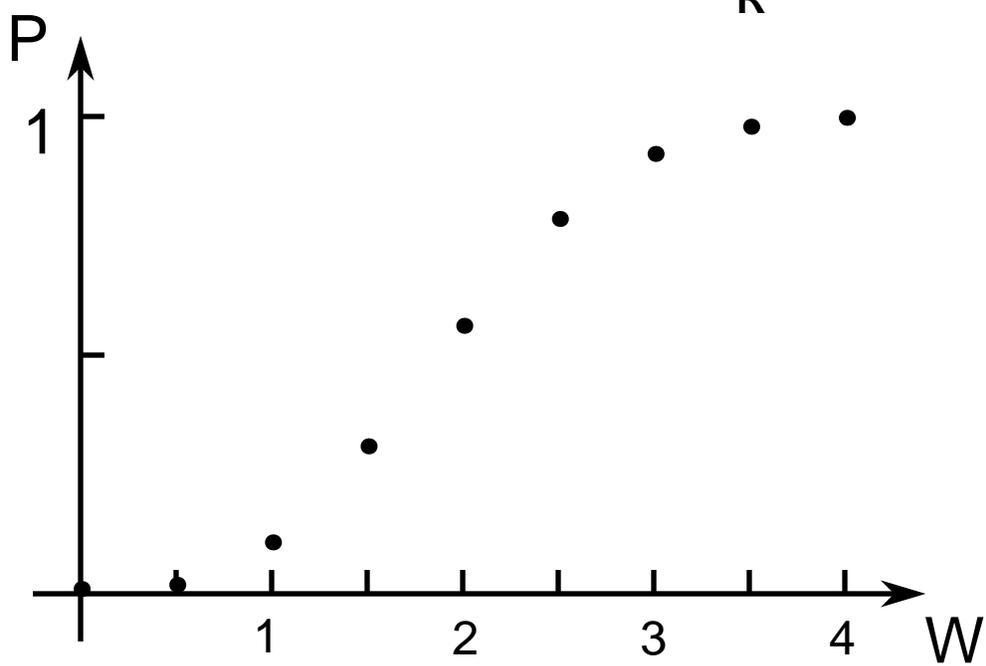
\end{document}